\documentclass[preprint,aps,prb,showkeys,showpacs]{revtex4}
\usepackage[dvips]{graphicx}
\DeclareGraphicsExtensions{.eps}

\usepackage{amsmath}

\begin{document}
\draft
\preprint{LA-UR-06-3305}
\title{Electrodynamics of a non-relativistic, non-equilibrium plasma}
\author{Shirish M. Chitanvis}
\affiliation{
Theoretical Division,\\
 Los Alamos National Laboratory,
 Los Alamos, New Mexico 87545}
\date{\today}
\begin{abstract}
A non-equilibrium plasma was studied using classical electrodynamic field theory. Non-linear interaction terms contribute to a finite lifetime for the dressed electrodynamic field. The lifetime exhibits a $\sim n^{-1} T_{e}^{3/2} T_{i}^{-2}T_{r}^{1/2}$ dependence, where $n$ is the number density, $T_{e}$ is the electron temperature, $T_{i}$ is the ion temperature, and $T_{r}$ is the temperature of the radiation field. The resulting width of the plasmon resonance is shown to decrease as equilibrium is approached. Dynamic screening leads to opaqueness of the plasma for low energy electromagnetic radiation.
This leads to a quadratic correction to the quartic Stefan-Boltzmann law. 
We also briefly discuss the effect of dynamic screening on fusion rates. 
Solitonic solutions to our non-linear wave equation allow localization of positive charges, which may enhance fusion rates.
\end{abstract}

\maketitle

\newpage
\section{Introduction}

Non-equilibrium plasmas have been investigated theoretically in several regimes by many authors\cite{spitzer,mihalas1,dharma98,murillo03}. These papers focus on approximations to deal with inter-particle collisions in plasmas. They develop various models to describe the temporal progress of a non-local-thermal-equilibrium (NLTE) plasma towards equilibrium.

Recently, probative experiments have ben performed on the passage towards equilibrium of a plasma in non-local thermal equilibrium (NLTE)\cite{benage06}. The paper by Taccetti at al\cite{benage06} gives a putative bound of about $1 ns$ as the time required for the NLTE plasma in their experiment to achieve equilibrium. There will be a similar bound for any other NLTE plasma. On time scales much smaller than this measure, one may safely assume that a NLTE plasma is frozen, viz., the electron temperature is different from the ion temperature. But light can travel appreciable distances on this time scale (e.g. $3cm$ in $0.1ns$), so that we need to consider the time dependence of electromagnetic fields. By extending the static methods of Jaffe and Brown\cite{brown00} and Chitanvis\cite{chitanvis06}, we can consider the classical statistics of the electrodynamic field in a non-relativistic NLTE plasma. By varying the electron temperature, the ion temperature, as well as the temperature of the electromagnetic field, we can study parametrically the passage of an NLTE towards local thermal equilibrium (LTE). Phenomena such as Bremsstrahlung  and inverse Compton scattering of hot electrons which equilibrate the temperature of electrons and the radiation field,  and electron-ion collisions, which equilibrate the energy between electrons and ions could be considered in the future to describe the temporal development of a non-equilibrium plasma towards LTE.

We have focused here on the dynamic dielectric constant of a non-equilibrium plasma, which can be measured in a laboratory. The dispersive properties of this quantity will be discussed. We propose that experiments be performed to test our predictions.

\section{Formulation of the statistical theory}

The electrostatic interaction between any two charges sitting in a collection of other charges is inevitably screened by intervening charges.  This effect becomes more dominant as the density of charges increases. Thus calculating the screening of the bare Coulomb interaction could be important for high-density plasmas. The paper by Varela et al\cite{varela} gives a recent overview of various methods brought to bear on this aspect of plasma physics. Some of these methods develop ideas based on the work of Fisher and Levin\cite{fisher93}. Our field-theoretic method bears a close resemblance to the approach of Brown and Jaffe\cite{brown00} and Chitanvis\cite{chitanvis06}.  We shall focus on the calculation of non-linear effects on the electrodynamics of a NLTE plasma. These non-linear effects are expected to dominate at high charge densities.

An assumption we will make is that on the scale that fluid motion takes place, the individual plasma components have separately reached thermal equilibrium, such that the ions of a specie possess a temperature which is different than the electron temperature.

The scalar potential $\phi$ of a plasma composed of a single atomic specie, having an atomic number $Z$, and number density $n$ is governed by:

\begin{eqnarray}
c^{-2}\partial_{t}^{2}\phi -\nabla^2 \phi &&= 4 \pi \rho \nonumber\\
\rho &&=\rho_{+}+\rho_{-}\nonumber\\
\rho_{+} &&=e ~ n~Z ~\exp(-Z e \phi/k_BT_{i}) \nonumber\\
\rho_{-}&&=- e~n~Z \exp(e \phi/kT_e)
\label{dh0}
\end{eqnarray}
where $c$ is the speed of light,
$\rho$ is the charge density,
where $k_B$ is the Boltzmann constant, 
$e$ is the electronic charge,
$T_{i}$ is the ion temperature,  $T_e$ is the electron temperature, and $n$ is the average number density of the plasma. Since we consider a non-relativistic plasma, we can neglect the effects of the vector potential, which are ${\cal O}(v/c)$, compared to the potential $\phi$.
We have assumed that positive and negative charge densities are governed by Boltzmann distributions, and the electron temperature is in general different from the ion temperature.
In so doing, we are making the assumption that quantum effects are unimportant.  Now, for the case of plasmas in local thermal equilibrium, Gruzinov and Bahcall\cite{gruzinov98}, among others, have estimated that quantum effects give rise to small deviations from Eqn.\ref{dh0} for the screening length. This could be important for certain applications.

Expanding the right hand side to third order in the electrostatic potential (basically in powers of a parameter $\Gamma$ which is analogous to the usual plasma expansion parameter), it can be shown that:
\begin{equation}
c^{-2}\partial_{t}^{2}\tilde \phi -\nabla^2 \tilde \phi \approx \left(\frac{Z  (Z \tau+1) }{\lambda_{DH}^2}\right)~(- \tilde \phi + \frac{\sigma_2 }{2!} \tilde \phi^2- \frac{\sigma_3}{3!} \tilde \phi^3 )
\label{dh01}
\end{equation}

where:
\begin{eqnarray}
\lambda_{DH} &&= \sqrt{\frac{k_B T_e}{4 \pi n_0 e^2}}\nonumber\\
\Lambda &&= \frac{\lambda_{DH}}{\sqrt{Z (Z \tau +1)}}\nonumber\\
\tilde \phi &&= \frac{\Lambda \phi}{e}\nonumber\\
\sigma_2&&= ~\Gamma~\frac{\tau^2~Z^2~ -1}{(Z \tau+1)}\nonumber\\
\sigma_3&&= ~\Gamma^{2}~\frac{\tau^3~ Z^3~ -1}{(Z \tau+1)}\nonumber\\
\Gamma &&= \frac{e^{2}}{\Lambda~k_{B}T_{e}}
\label{dh1}
\end{eqnarray}
where $\tau = T_e/T_{i}$, $\Gamma$ is a parameter analogous to the usual plasma parameter. Basically, we use $\Lambda$ as a length scale rather than the mean free distance, since $\Lambda$ appears naturally in the theory.
$n_0$ is the average number density of the plasma, and $\lambda_{DH}$ is the standard Debye-Huckel screening length, which holds in the case of a single-component plasma.

Equation \ref{dh1} demonstrates as long as $\Gamma << 1$ the coupling constants $\sigma_2 < 1,~\sigma_3<1$, as long as $Z \tau$ not too high.  In this sense, we have tried to ensure that our series expansion is at least formally convergent.
The value of $\tau$ ranges between a number much greater than one to unity, as the non-equilibrium plasma tends towards equilibrium.  
As an example, we note that at $T_{e} = 1 keV$, $\Lambda \sim 10^{-9}cm$, $\Gamma \sim 0.06$.  Hence, our expansion is valid for $Z \tau \le 1/0.06 \approx 16.7$. For $Z \sim 1$, our expansion is applicable to plasmas fairly out of equilibrium.

 Furthermore, note that retaining non-linear terms on the right hand side of Eqn.\ref{dh0}, in powers of $\Gamma$, is equivalent to seeking corrections due to the high density of the plasma\cite{murillo03}. The number of terms that need to be retained thus depend on $\Gamma$.

Now in the linear case when $\sigma_2=\sigma_3=0$ in Eqn.\ref{dh01}, it is easy to see that our plasma has an effective screening length $\Lambda$ which is given by the standard single-component Debye-Huckel length {\em divided} by a factor of $\sqrt {(Z \tau+1) Z}$. Thus this correction can be significant for large $Z$, and for cases when there is a large temperature difference between the electrons and ions.

Note that Eqn.\ref{dh1} for the electrostatic potential can be obtained by extremizing the following Lagrangian density:

\begin{eqnarray}
{\cal L}(\tilde \phi) &&= \frac{1}{2}((\partial_{t'} \tilde \phi)^2- |\vec \nabla' \tilde \phi|^2) -V(\tilde \phi)\nonumber\\
V(\tilde \phi) &&=  (z\tilde \phi +\frac{1}{2}\tilde \phi^2  - \frac{\sigma_2}{3!}~\tilde \phi^3 + \frac{\sigma_3}{4!} \tilde \phi^4)
\label{dh3}
\end{eqnarray}
wher $t'=c t/\Lambda$,
where the prime on the gradient operator indicates we are using $\Lambda$ as a length scale. 

The cubic term can be interpreted as a three-wave interaction, and the quartic a four-wave process-- If second quantization is employed, these terms would correspond to three-photon and four-photon processes. One may perhaps expect these terms to be important in the high energy density regime.  Moreover, the theory bears a formal resemblance to the theory of phase transitions. However, we do not have a phase transition to contend with, and hence the effect of fluctuations could be minimal, and mean field theory (Eqn.\ref{dh1} corresponds to  a dynamic Debye-Huckel approximation) would be sufficient to describe the system. In fact detailed calculations presented in this paper show that higher order processes are not be important as far as renormalization of the plasmon frequency are concerned.  But they are essential if one wishes to consider the lifetime of the dressed electromagnetic field.

Since we have a time-independent Lagrangian density, the momentum $\pi \equiv \partial {\cal L}/\partial \dot {\tilde \phi} =\partial_{t'} \tilde \phi$. Hence the Hamiltonian density ( i.e. the energy density) is:

\begin{equation}
{\cal H} =\pi {\dot {\tilde \phi}} -{\cal L} = \frac{1}{2}( (\partial_{t'} \tilde \phi)^2+|\vec \nabla' \tilde \phi|^2) + V(\tilde \phi)
\label{dh3a}
\end{equation}
Note that the Hamiltonian density (energy) is positive semi-definite.
We define the following partition function in order to study the statistical mechanics of the system described by this energy density:

\begin{equation}
{\cal Z} = \int {\cal D} \tilde \phi \exp(-\int d^4 x' ~{\cal H}(\tilde \phi)/(k_{B}T_{r}))
\label{dh4}
\end{equation}
where $T_{r}/T_{e}$ is the temperature of the electromagnetic field in units of the electron temperature, since we are using $k_{B}T_{e}$ as the unit of energy.
 The functional integral over all fields denotes a sum over all thermodynamic states. In this sense, we take into account fluctuations around Eqn. \ref{dh1}. We consider Eqn. \ref{dh1} a mean-field approximation, and a non-linear, dynamic extension of the standard Debye-Huckel equation.
  
In order to simplify calculations, we will define $\Phi = (\sqrt{T_{e}/T_{r}})~\tilde \phi$,
and $\sigma_{2}\to \sigma_{2}'=\sqrt{T_{r}/T_{e}}~\sigma_{2}$, and $\sigma_{3}\to\sigma_{3}'=(T_{r}/T_{e})~\sigma_{3}$.
Since we generally expect $T_{e}>>T_{r}$ for an NLTE plasma, $\sigma_{3}' << \sigma_{2}'$.

In terms of $\Phi$, the partition function can be written as:

\begin{equation}
{\cal Z} = \int {\cal D} \Phi \exp(-\int d^4 x' ~{\cal H}(\Phi))
\label{dh5a}
\end{equation}

We can now use standard techniques from (Euclidean) field theory\cite{ramond} to obtain the lowest order non-trivial corrections to the self-energy, or equivalently, the screening length.  The correction to ${\cal O}(\sigma_2'^2)$ is obtained by noting that when higher order corrections due to the non-linear term in $V(\Phi)$ is taken into account:

\begin{equation}
\hat G_0(\omega,k) \to \left(\hat G_0^{-1}(\omega,k) - \hat \Sigma(\omega,k)\right)^{-1}
\label{dh4a}
\end{equation}

where $\hat G_0(\omega,k)$ denotes the momentum-space inverse of the operator $(-\partial_{t'}^{2}-\nabla^2 + 1)$ in the units indicated above, viz., $\hat G_0(\omega,k)=(\omega^{2}+k^{2}+1)^{-1}$, and where $\hat \Sigma(\omega,k)$ is the self-energy. Note that this is somewhat different than the usual Klein-Gordon-like operator in momentum-space, $(\omega^{2}-k^{2}-1)^{-1}$. One can obtain an equivalence by letting $k \to i k$ and $m \to i m$.

We used Mathematica to generate symbolically the second order contribution (the first non-trivial correction) from the cubic term in the energy functional.
The net result is:

\begin{eqnarray}
\hat \Sigma_{polarization-like}(\omega,p) &&= \frac{\sigma_2'^2}{48} \int \frac{d\Omega~d^3k}{(2 \pi)^4} \hat G_0 (\Omega,\vec k) \hat G_0(\omega-\Omega,\vec p - \vec k) \nonumber\\
\hat G_0(\omega,\vec p) &&= \frac{1}{\omega^{2}+p^{2} + 1}
\label{dh5}
\end{eqnarray}

The form of the integral can be evaluated using dimensional regularization\cite{ramond}:

\begin{eqnarray}
&&\hat \Sigma(\omega,k)_{polarization-like} = \nonumber\\
&&\frac{{\sigma_2 '}^2}{48 \cdot 16 \pi^{2}}~( 2-\gamma +2~i~\pi+ 2 \ln(4 \pi) -
\sqrt{1 - 4/(\omega^{2}+k^{2})}\ln\left(\frac{\sqrt{1 - 4/(\omega^{2}+k^{2})}+1}{\sqrt{1 - 4/(\omega^{2}+k^{2})}-1}  \right) 
\label{dh5a}
\end{eqnarray}

We will work in the regime where $T_{e}>> T_{r}$, so that the quartic term in $V(\Phi)$ can be safely ignored, using the scaling arguments described earlier.

The appearance of an imaginary term preceding the logarithm gives rise to a lifetime for electromagnetic excitations, analogous to the lifetime of a quasi-particle. Furthermore, there is an additional contribution to the lifetime for $1 < 4/(\omega^{2}+k^{2})$.

The net effect is that we account to ${\cal O}(\sigma_{2}'^{2})$ the screening effects on the pure Coulomb potential. It is straightforward to obtain the effective wave-equation, obtained in lieu of Eqn.\ref{dh3}:

\begin{equation}
{\cal L}  \to {\cal L}_{effective} \equiv \frac{1}{2}((\partial_{t'} \Phi)^2- |\vec \nabla' \Phi|^2 -\Phi|^2+ \int d^{4}x' \Phi(x')~\Sigma_{polarization-like}(x'-x)\Phi(x))
\label{Leff}
\end{equation}

The effective wave equation follows simply, by extremizing this Lagrangian, and the spectral form of its Green's function can be obtained as follows:
 
\begin{equation}
\hat {\cal G}(\omega,k) = \left(\omega^{2}-k^{2}-1+\hat \Sigma_{polarization-like}(\omega,k))\right)^{-1}
\label{die1}
\end{equation}

\begin{figure}[htbp]
\begin{center}
\includegraphics{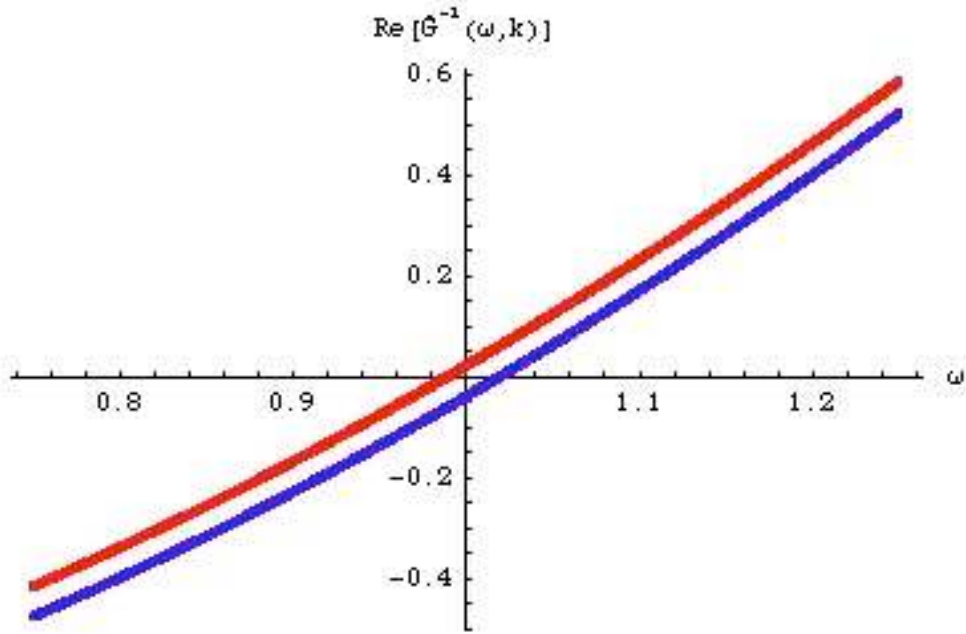}
\caption{The blue curve corresponds to $\sigma_{2}'=1$ and has a zero at the plasmon frequency of $\omega_{p}\approx1$ (in the units chosen), while the red curve corresponds to $\sigma_{2}'=10$ and also has a zero at $\omega_{p}\approx 1$.  This shows that the plasmon frequency $\omega_{p}$ does not depend sensitively on the coupling constants, which in turn depend on the plasma parameters.  Note that while the theory itself is formally valid for values of the coupling constant less than one, we have used larger values of the coupling constants only for exposition.}
\label{FIG1}
\end{center}
\end{figure}

Basically, Eqn. \ref{die1} yields a plasmon resonance at :

\begin{equation}
\omega=\pm\sqrt{c^{2 }k^{2}+\omega_{p}^{2}}
\label{pl1}
\end{equation}
where $c=1$ in the units we are using.
This is a perturbative result, valid for $\sigma_{2}' < 1$.  
One might be tempted to assume from Eqns. \ref{dh5a} that the overall contribution is $<<1$, given the small numerical pre-factor.  As speculated earlier, this is indeed the case, as far as the plasmon frequency is concerned--
In the units chosen, $\Lambda=1$, and  $\omega_{p}^{2}\approx c^{2}/\Lambda^{2} +{\cal O}(\sigma_{2}'^{2})\approx 1$. 
The effects of interaction on the renormalization of the plasmon frequency can be gauged by studying the zero of $Re({\cal G}^{-1}(\omega,k=0.1))$ as displayed in Fig. \ref{FIG1}. The zero occurs at the plasmon frequency $\omega_{p}$.  The plasmon frequency is insensitive to the coupling constants, which in turn depend on the parameters of the plasma.

\begin{figure}[htbp]
\begin{center}
\includegraphics{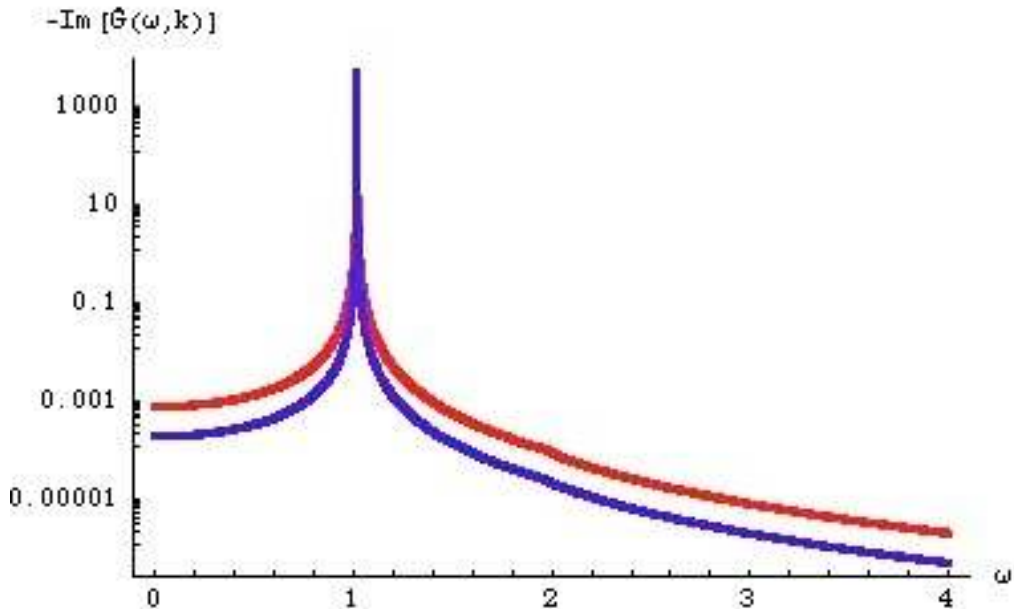}
\caption{This plot of the spectral weight. It clearly shows the plasmon structure, broadened by three-wave interactions. The blue curve corresponds to $\sigma_{2}'=1/2$ and refers to a case closer to equilibrium, and the red curve corresponds to an NLTE  case of $\sigma_{2}'=1$.}
\label{FIG2}
\end{center}
\end{figure}

The negative of the imaginary part of the structure factor $\hat {\cal G}(\omega,k)$ yields the spectral weight.  The spectral weight manifests the plasmon resonance.  This quantity should be accessible to experiments.
Qualitatively, this is what one expects to occur in a plasma\cite{ichimaru}. For sufficiently large frequencies, far from the plasmon frequency, one expects a long-lived wave.
This effect is exemplified in Fig.\ref{FIG2}. This figure shows the spectral weight for two different values of the coupling constant. The red curve refers to $\sigma_{2}'=1$ and signifies a case more out of equilibrium, and the blue curve with $\sigma_{2}'=1/2$ refers to case closer to equilibrium. The spectral weight is sensitive to the value of the coupling constants (see Figure \ref{FIG2}) over part of the frequency spectrum. Notice also that the width narrows as the system approached equilibrium.

\section{Dynamic screening effects}

The dielectric constant $\epsilon(\omega,k)$ of the plasma may be obtained via the following formal identification:

\begin{eqnarray}
\hat {\cal G}(\omega,k) &&= \left(\omega^{2}-k^{2}-1+\hat \Sigma_{polarization-like}(\omega,k))\right)^{-1}\nonumber\\
&&=\left(\epsilon(\omega,k) \omega^{2} - k^{2}\right)^{-1}\nonumber\\
\epsilon(\omega,k) &&= 1 -\frac{1 - \hat \Sigma_{polarization-like}(\omega,k)}{\omega^{2}}
\label{dieform}
\end{eqnarray}

\begin{figure}[htbp]
\begin{center}
\includegraphics{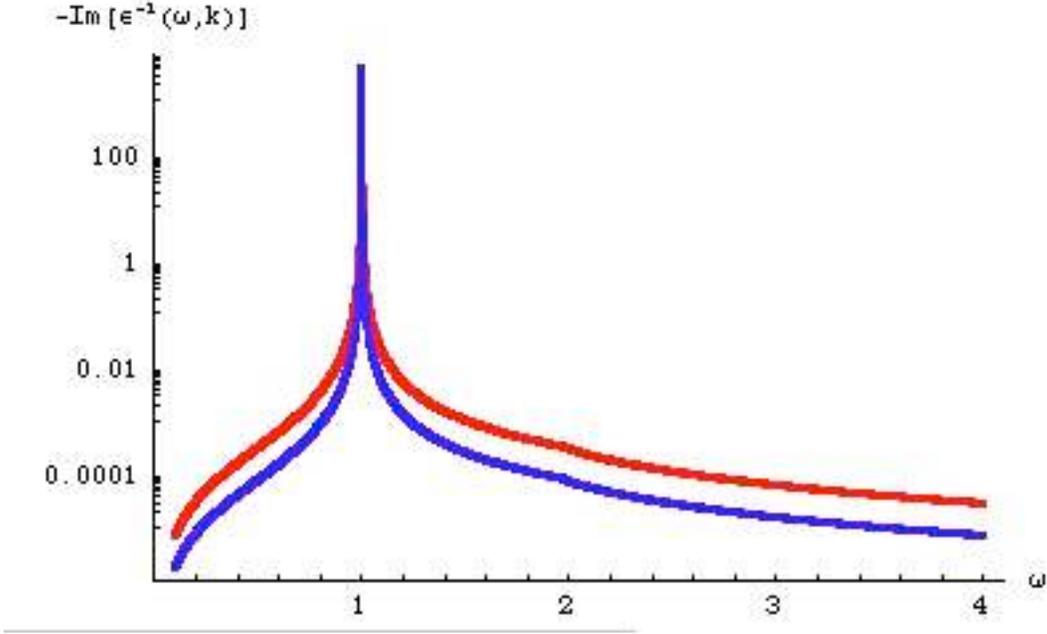}
\caption{This plot of the negative of the imaginary part of the inverse of the dielectric constant, which is proportional to the structure factor. It clearly shows the plasmon structure, broadened by three-wave interactions. The blue curve corresponds to $\sigma_{2}'=1/2$ and refers to a case closer to equilibrium, and the red curve corresponds to an NLTE  case of $\sigma_{2}'=1$.}
\label{FIG3}
\end{center}
\end{figure}

We have plotted in Fig. \ref{FIG3} the negative of the imaginary part of the inverse of the dielectric constant, which is proportional to the structure factor ${\cal S}(\omega, k)$\cite{ichimaru} of the NLTE plasma. The structure factor may be probed in future experiments.

\section{Relaxation Rates}

Third-order interactions (corresponding to three photon processes) lead to finite lifetimes for electromagnetic excitations. Jaffe and Brown\cite{brown00} have pointed out in the static case, there is a correspondence between our approach, which focuses on the electromagnetic field, and the ensemble consisting of electrons and ions. A similar connection must exist for the dynamic case considered here as well.  Relaxation rates for collisions between ions and electrons can be computed for various regimes\cite{dharma01}. Not only does our theory account for such inter-particle collisions in an integrated sense\cite{brown00}, but addresses radiative processes as well. The inverse of the rate we consider here is a measure of the time required for electromagnetic energy in a specific part of the spectrum to be given to other parts of the spectrum.

To the first order in the coupling constant $\sigma_{2}'$, the relaxation rate, written as an inverse lifetime is:

\begin{eqnarray}
\tau_{em}^{-1} &&= \left(\frac{c}{\Lambda}\right)~\left(\frac{\sigma_{2}'}{32 \pi}\right)~\sqrt{2 \pi (T_{r}/T_{e})} \nonumber\\
&& =\left(\frac{c}{\Lambda}\right)~\left(\frac{\Gamma (\tau^2~Z^2~ -1)}{32 \pi (Z \tau+1)}\right) ~\sqrt{2 \pi(T_{r}/T_{e})}
\label{rel1}
\end{eqnarray}
for $1 > 4/(\omega^{2}+k^{2})$. Here we have re-introduced dimensionful parameterization.  

For the case of extreme non-equilibrium, for $Z>>1$, when $\tau >>1$, we can extract the main functional dependencies as:

\begin{equation}
\tau_{em}^{-1} \sim~n^{-1}~T_{e}^{3/2}~T_{i}^{-2}~T_{r}^{1/2}
\label{rel2}
\end{equation}

This result can be compared to the usual Spitzer rate (two-temperature plasma), which is restricted to electron-ion collisions, and yields a $T_{e}^{3/2}$ temperature dependence.

For $1 < 4/(\omega^{2}+k^{2})$, we obtain an additional contribution to the lifetime which depends on the region of the spectrum being probed:

\begin{eqnarray}
\tau_{em}^{-1} && =\left(\frac{c}{\Lambda}\right)~\left(\frac{\Gamma (\tau^2~Z^2~ -1)}{32 \pi (Z \tau+1)}\right) ~\sqrt{(T_{r}/T_{e})}\cdot\nonumber\\
&&\sqrt{\left( 2 \pi - {\cal I}m\left(\sqrt{1 - 4/(\omega^{2}+k^{2})}\ln\left(\frac{\sqrt{1 - 4/(\omega^{2}+k^{2})}+1}{\sqrt{1 - 4/(\omega^{2}+k^{2})}-1}  \right)  \right)\right)}
\label{rel2}
\end{eqnarray}

\section{Radiative properties of a NLTE plasma}

We have found that the dispersion relation is given approximately $\omega=\sqrt{k^{2}+\omega_{p}^{2}}$.
This implies that electromagnetic waves not satisfying this relation exist at best in an evanescent fashion in the plasma.
One expects on physical grounds that this absorptive-transparency phenomenon will affect the radiative properties of a black-body in which a plasma exists as well. 

One can estimate this effect by modifying the standard calculation of the energy density of the electromagnetic field by noting that the density of states $g(\omega)$ per unit volume for the present case is:

\begin{eqnarray}
\tilde g(k) dk &&= \frac{2 \cdot 4 \pi \cdot k^{2}}{(2 \pi)^{3}}~dk \nonumber\\
g(\omega) d\omega &&= \frac{\omega~\sqrt{\omega^{2} - \omega_{p}^{2}}}{c}~d\omega~\forall \omega \ge \omega_{p}
\label{bb1}
\end{eqnarray}
Here we have restored dimensionality of all parameters.
The resulting energy radiation flowing per unit area, per unit time, to leading order in $\omega_{p}$ is then:

\begin{equation}
S(T_{r},\omega_{p}) = \sigma_{SB} T_{r}^{4}~\left(1-\frac{5 \hbar^{2} \omega_{p}^{2}}{4 \pi^{2}k_{B}^{2}T_{r}^{2}}+...\right)
\label{bb2}
\end{equation}
where $\sigma_{SB}$ is the Stefan-Boltzmann constant.

We note the following estimates: 

$\bullet$ The plasmon frequency $\omega_{p}^{0}\sim10^{16}Hz$ at an electron temperature of $1 keV$, and a number density $\approx 10^{18}cm^{-3}$.  This corresponds to photon energy $\hbar \omega_{p}^{0}\sim 10 eV$. 
This suggests that our corrections to the Stefan-Boltzmann law can be significant for a radiation field temperature $T_{r}\sim 10 eV$.

\section{Quantum tunneling}

We will consider briefly in this section the effect of dynamic screening on fusion rates.  Toward the end of the section, we indulge in speculating how the thermal rate of fusion may be affected in NLTE plasmas. This may be considered a prelude to a full-fledged study of nuclear fusion rates in non-equilibrium plasmas.
We have restored dimensionality of all parameters in this section.

Now the WKB penetration factor ${\cal P}(E)$ for charged nuclear particles interacting via a pure Coulomb term at sufficiently large distances compared to the nuclear lengths is: 

\begin{eqnarray}
{\cal P}(E) &&= \exp\left(-2 \int_{r_{1}}^{r_{2}}\kappa(r)dr\right) \nonumber\\
\int_{r_{1}}^{r_{2}}\kappa(r)dr &&= \frac{\pi Z Z' e^{2}}{\hbar v} \left(1-\frac{2}{\pi}~\sin^{-1}\gamma^{-1/2} \right) -\frac{\mu v R} {\hbar} ~\sqrt{\gamma-1}\nonumber\\
\gamma &&= \frac{Z Z' e^{2}}{E R}\nonumber\\
E &&= E_{k}+ \frac{Z Z'e^{2}\omega_{p}}{c}\nonumber\\
E_{k}&&=\frac{\mu v^{2}}{2} 
\label{pf1}
\end{eqnarray}

where $r_{1}=R$ is the turning point where nuclear attractive forces begin to dominate Coulomb repulsion, and the other turning point $r_{2}=Z Z'e^{2}/E$, $Ze$ and $Z'e$ being the charges on the interacting particles. $\mu$ is the reduced mass of the two nuclear interacting particles, $v$ is the relative speed, and $E$ is the sum of the kinetic energy and the static screening effect.

Now, from Eqn.\ref{dh5a}, the pure Coulomb interaction is modified thusly:

\begin{equation}
\frac{1}{r} \to \frac{\exp(-r~\sqrt{\omega_{p}^{2}/c^{2}-E_{k}^{2}/(c^{2}\hbar^{2})}) }{r}
\label{pf2}
\end{equation}
where for sufficiently low values of the kinetic energy $E_{k}$, the radical in Eqn.\ref{pf2} remains real. It is seen that accounting for the finite speed of light causes the screening length to increase, i.e., the range of the interaction increases, allowing for repulsion over a larger distance.  Hence we expect the penetration factor to decrease.

Upon expanding in powers of $r$, since we are only interested in short separations, and further, retaining only leading order terms in $E_{k}$, we obtain an addition to the argument in the penetration factor:

\begin{eqnarray}
\int_{r_{1}}^{r_{2}}\kappa(r)dr &&\to \int_{r_{1}}^{r_{2}}\kappa(r)dr + \Delta \nonumber\\
\Delta &&=\left(\frac{E_{k} \alpha \gamma^{2}}{2 c \hbar^{2}\omega_{p}\sqrt{\gamma-1}}\right)~ \left( \frac{\alpha }{\hbar v \gamma} - \frac{\pi R \mu v}{2 \hbar}\right)\nonumber\\
\alpha&&=Z Z' e^{2}
\label{pf3}
\end{eqnarray}

Hence, by accounting for the finite speed of light, we have shown that retardation effects will affect the fusion rate. Salpeter\cite{salpeter54} referred to dynamic effects in passing in his seminal paper, while more recently Shaviv and Shaviv\cite{ss00}, and Bahcall et al\cite{bahcall02} have considered such effects in more detail. Here we have presented an alternative treatment of the same effect.

Upon utilizing $\omega_{p}/c \sim 10^{9}cm^{-1}$, $E_{k}\sim 10 keV$, $R\sim 10^{-13}cm$, $\mu \sim 10^{-25}g$, we find that the first term in Eqn.\ref{pf3} is greater than the first, so that the net effect for this set of parameters is to decrease the fusion rate, albeit by a small amount ${\cal O}(10^{-4})$, compared to Eqn.\ref{pf1}, which yields a contribution of $\sim {\cal O}(10^{-1})$.

Note that this theory is based on the assumption that the charge density follows the classical Boltzmann distribution even at nuclear distances. This point has been questioned by many authors, e.g. Gruzinov and Bahcall\cite{gruzinov98} for the case of a plasma in local thermal equilibrium. They find that quantum effects change the screening length by approximately $1\%$ in the solar interior. We will consider such effects for a NLTE plasma in a future paper.

It is well-known that the fusion rate at a finite temperature is directly proportional to the relative fraction of the reactants.  Hence it is possible that if concentration of reactants is increased locally, it would enhance fusion in that spatial region.  To see that such charge configurations may occur in our system, let us consider the wave equation to the second order in the potential:

\begin{equation}
\left(\partial_{t}^{2}- \nabla^{2}\right) \Phi = - \Phi + \frac{\sigma_{2}'}{2} \Phi^{2}
\label{wv1}
\end{equation}

We seek a traveling wave solution in one dimension of the form $\Phi(x-ut)$, where the wave-speed $u$ is measured in units of the speed of light.  Redefining $\xi=(x-ut)/\sqrt{1-u^{2}}$, with $u < 1$, we see that for the special case of $z=0$, the solution is:

\begin{equation}
\Phi(\xi) = \left(\frac{3}{ \sigma_{2}'}\right)~{\rm sech}^{2}(\xi/2)
\label{wv2}
\end{equation}

\begin{figure}[htbp]
\begin{center}
\includegraphics{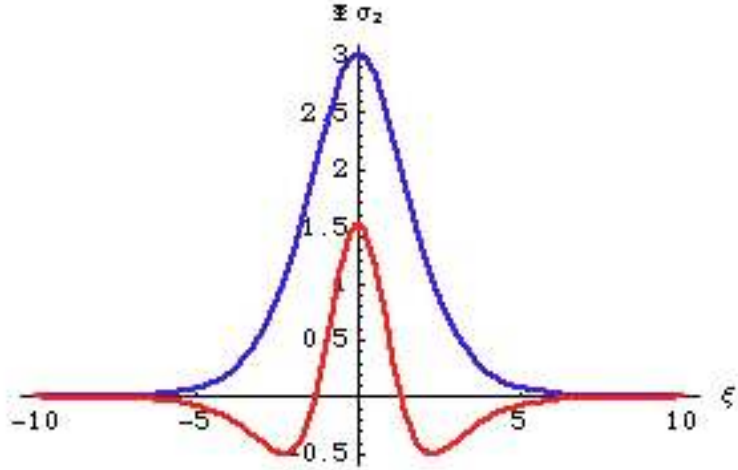}
\caption{The blue curve is a solitonic solution to Eqn.\ref{wv1}. The red curve is the corresponding charge density.  Note how non-linear effects serve to localize the positive charge.}
\label{FIG4}
\end{center}
\end{figure}

This is of course a traveling soliton.  It is interesting to see what charge distribution is carried along with this soliton. This can be estimated by computing the curvature of the solution, viz., $-\partial_{\xi}^{2}\Phi(\xi)$.  We illustrate this is Fig.\ref{FIG4}.  Notice that the blue curve is the potential, which shows a change in the curvature as we move away from the origin.  This is manifested in the charge density, displayed as the red curve.  Notice how the positive charge gets localized by a surrounding negative charge density. Furthermore, we have lowered in Fig. \ref{FIG4}the charge density by a scaling factor of $\sigma_{2}' < 1$. This represents a deviation from the average charge density, which will be much lower.  The fact that positive charges (nuclei) have been localized will enhance the rate of fusion.  Of course, whether such localization occurs in a given plasma depends on initial and boundary conditions on the NLTE plasma.

\section{Conclusions}
We studied parametrically dynamic screening effects in a three-temperature plasma. A statistical theory of the electromagnetic field interacting with charges was developed for this purpose. Diagrammatic techniques were utilized to obtain leading order estimates of the lifetime of the plasmon resonance due to three-wave interactions. Effects of dynamic screening on radiative properties were deduced.  We speculated on the manner in which solitonic solutions could allow localized enhancement of positive charges, which in turn would increase fusion rates.

We propose that the techniques developed, and results obtained in this paper could be tested in future experiments.

\bibliographystyle{apsrev.bst}

\bibliography{pre_scc_bibdata}
\newpage

\end{document}